\documentclass[conference]{IEEEtran}

\usepackage{amsmath,amsthm,amssymb,paralist,subfigure,cite,graphicx,color,amsbsy,float}
\usepackage{pifont}
\usepackage{stmaryrd,mathrsfs,url}
\usepackage[table]{xcolor}
\usepackage{cuted,mathtools,lipsum}
\usepackage[ruled,vlined,linesnumbered]{algorithm2e}
\usepackage{soul}
\usepackage{microtype} 
\usepackage{graphicx}
\usepackage{hyperref}
\hypersetup{
	colorlinks=true,
	linkcolor=cyan,
	filecolor=mnodea,
	urlcolor=cyan,
	citecolor=lime,
}

\newtheorem{rem}{Remark}

\def\mb{\mathbf}

\def\mc{\mathcal}

\IEEEoverridecommandlockouts
\begin{document}
\title{\huge \bf
Fully Distributed and Quantized Algorithm for MPC-based Autonomous Vehicle Platooning Optimization}

\author{\IEEEauthorblockN{ Mohammadreza Doostmohammadian}
\IEEEauthorblockA{
\textit{Faculty of Mechanical Engineering}, \\
\textit{Semnan University}, Iran, \\ \texttt{doost@semnan.ac.ir}}
\and
\IEEEauthorblockN{Alireza Aghasi}
\IEEEauthorblockA{\textit{Electrical Engineering Department}, \\ \textit{ Oregon State University},
USA, \\
\texttt{alireza.aghasi@oregonstate.edu}}
\and
\IEEEauthorblockN{Hamid R. Rabiee}
\IEEEauthorblockA{\textit{Computer Engineering Department}, \\ \textit{ Sharif University of Technology},
	Iran, \\
	\texttt{rabiee@sharif.edu}}
}

\maketitle

\begin{abstract}
	Intelligent transportation systems have recently emerged to address the growing interest for safer, more efficient, and sustainable transportation solutions. In this direction, this paper presents distributed algorithms for control and optimization over vehicular networks. First, we formulate the autonomous vehicle platooning framework based on model-predictive-control (MPC) strategies and present its objective optimization as a cooperative quadratic cost function. Then, we propose a distributed algorithm to locally optimize this objective at every vehicle subject to data quantization over the communication network of vehicles. In contrast to most existing literature that assumes ideal communication channels, log-scale data quantization over the network is addressed in this work, which is more realistic and practical. In particular, we show by simulation that the proposed log-quantized algorithm reaches optimal convergence with less residual and optimality gap. This outperforms the existing literature considering uniform quantization which leads to a large optimality gap and residual.       
		
\end{abstract}

\section{Introduction} \label{sec_intro}
Emerging notions of cloud-computing and multi-agent processing motivate distributed and parallelized algorithms in large-scale applications from filtering and estimation over sensor networks \cite{spl18,zamani2016iterative,zayyani2023adaptive} to distributed machine learning \cite{qureshi2022stochastic,csl2021,xin2020decentralized} and computing resource allocation \cite{scl23,yuan2020profit}. Distributed algorithms for control and optimization advance centralized solutions in terms of scalability, robustness to central-node-of-failure, dynamic flexibility for node addition/removal, and enabling parallel processing capabilities. In this direction, the rapid advancement of intelligent transportation systems (ITS) requires innovative distributed solutions to enhance the safety and sustainability of vehicular networks \cite{dimitrakopoulos2010intelligent,singh2015recent}, particularly in the context of multi-agent systems such as autonomous vehicle platooning \cite{li2022cooperative,badnava2021platoon}. In decentralized ITS, each vehicle can independently process local information and communicate with its neighbours, leading to improved adaptability to dynamic traffic conditions \cite{hamdipoor2023safe} and reduced latency in control decision-making \cite{liang2021distributed}. 

Different distributed algorithms for distributed control and optimization over ITS are proposed in the literature. Consensus algorithms \cite{cortes2008distributed,lcss21} are fundamental in distributed control, enabling multiple agents (vehicles) to reach an agreement on a particular state or decision. These algorithms are particularly useful in scenarios where vehicles need to synchronize their actions, such as maintaining safe distances in platooning \cite{han2022distributed}. Distributed MPC is an extension of traditional MPC that allows multiple agents to optimize their control inputs while considering the dynamics of neighbouring vehicles. This approach decomposes the global optimization problem into smaller subproblems that can be solved locally \cite{shen2022fully,shen2022nonconvex,ferramosca2013cooperative}. Techniques such as dual decomposition \cite{gao2023distributionally} and alternating direction method of multipliers (ADMM) \cite{zhou2023distributed,cdc22} have been employed to facilitate coordination among vehicles while ensuring optimal performance. Similarly, distributed optimization algorithms aim to solve optimization problems across a network of agents. Techniques such as gradient descent and subgradient methods have been adapted for distributed settings, allowing vehicles to collaboratively optimize objectives like route planning \cite{camponogara2024distributed} or energy consumption \cite{liu2023hierarchical}. These algorithms typically involve local computations and limited communication, making them suitable for real-time applications in ITS \cite{zhao2022online}. In more unknown and dynamic setups, vehicles learn optimal policies through interactions with their environment, using techniques like multi-agent reinforcement learning that enable vehicles to adapt to changing traffic conditions \cite{zhou2022multi,berbar2022reinforcement}. 

In this paper, we explore fully distributed and quantized algorithms for MPC in platooning optimization. Such algorithms not only align with the principles of cooperative driving but also pave the way for more resilient and efficient transportation networks by addressing real-world constraints such as data quantization over the information-sharing network of vehicles. Vehicles continuously generate a vast amount of data from various sensors and systems. By quantizing this data, the volume of information transmitted between vehicles can be significantly reduced. This reduction minimizes the bandwidth requirements for communication and ensures that the available network resources are utilized effectively. This notion of quantized data exchange and resource-constrained data sharing is not considered in most existing literature. In this direction, this paper proposes distributed optimization techniques under log-scale quantization setups. In many distributed optimization scenarios, the data being processed can vary significantly in magnitude. For instance, in vehicle platooning, the distances between vehicles, speeds, and other parameters can span several orders of magnitude. Log-scale quantization allows for more precise representation of such a dynamic range of values. By ensuring that smaller values are represented with higher fidelity, log-quantized distributed optimization leads to more accurate optimization performance. This idea is used, for example, in log-quantized distributed resource allocation \cite{ecc22}. In particular, we show by simulation that the log-quantized optimization as compared to the uniform quantization \cite{pu2016quantization,rikos2024distributed,yi2014quantized} results in a lower optimality gap. This is because log-quantization allocates more bits to represent smaller (near-optimal) values and fewer bits for larger values; particularly the gradients or optimization updates have more accurate values near the optimal point. 

The rest of the paper is organized as follows. We first model the MPC framework for vehicle platooning as a distributed optimization problem in Section~\ref{sec_prob}. In Section~\ref{sec_dist}, we propose a distributed log-quantized gradient-tracking-based algorithm to solve this optimization problem. We validate our algorithm by simulation in Section~\ref{sec_sim} and particularly show that the log-quantized technique outperforms uniform quantization. Finally, Section~\ref{sec_con} concludes the paper.  

\section{The MPC Framework for Platooning} \label{sec_prob}
\begin{figure}[hbpt!]
	\centering
	\includegraphics[width=3.5in]{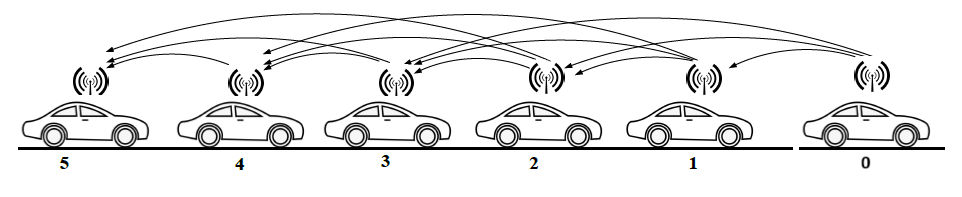}
	\caption{This figure shows the platoon of $n$ AVs and their connecting communication network. Every AV is modelled by linear dynamics, which along with the communication network defines the overall platoon dynamics. }
	\label{fig_platoon}
\end{figure}

We consider a platoon of $n$ connected autonomous vehicles (as in Fig.~\ref{fig_platoon}) communicating over a network $\mc{G}_W = \{\mc{V},\mc{E}\}$ with $\mc{V}$ denoting the set of vehicles as nodes and $\mc{E}$ as the set of communication links among the vehicles. The sequence of AVs follows a leader with controlled velocity and acceleration in a platooning setup.  
A link $(i,j) \in \mc{E}$ implies a data-sharing link between two nearby AVs $i,j$. The set of neighboring vehicles of AV $i$ is defined as $\mc{N}_i = \{j|(j,i) \in \mc{E}\}$.
Each vehicle $i$ is modelled by linear double-integrator dynamics as follows:
\begin{align} 
	p_i(k+1) &= p_i(k) + \tau v_i(k) + \frac{\tau^2}{2} u_i(k) \\
	v_i(k+1) &= v_i(k) + \tau u_i(k)
\end{align}
with $\tau$ as the sampling time, $k$ as the time-index, $u_i$ as the input, and states $p_i,v_i,a_i$ as the position, velocity, and acceleration of the AV $i$. This platooning model is constrained with some state and control constraints as 
\begin{align} \label{eq_constv}
	v_{\min} &\leq v_i(k) \leq v_{\max} \\ \label{eq_consta}
	a_{\min} &\leq u_i(k) \leq a_{\max} \\ \label{eq_constp}
	p_{i+1}(k)-p_{i}(k) &\geq l + \epsilon v_i(k) - \frac{(v_i(k) - v_{\min})^2}{2 a_{\min}}
\end{align} 
with $l$ as the vehicle length and $\epsilon$ as the reaction time of the AV dynamics.

A model predictive control (MPC) model is used for control of the AV platooning at time-instant $k$ \cite{shen2022fully,shen2022nonconvex}. For notation simplicity, we drop the dependence on $k$ from this point onward in the paper unless where it is needed. Let $\delta$ be the desired constant space between two nearby AVs in the platoon. We define three error parameter vectors: $e_p(k) := (p_0 - p_1 - \delta, \dots, p_{n-1} - p_n - \delta)$ as the relative (spacing) position error, $e_v(k) := (v_0 - v_1, \dots, v_{n-1} - v_n)$ as the relative velocity error, and $e_u(k) := (u_1 - u_2, \dots, u_{n-1} - u_n)$ as the relative input error between every two adjacent AVs. The terms $p_0,v_0,u_0$ define the state of the leader. 
Using the error terms, one can write the vector $u=(u_1,\dots,u_n)$ as $u = -S e_u + u_0 \mb{1}_n$ with $\mb{1}_n$ as the column vector of all ones of size $n$ and
 \begin{align} \label{eq_S}
 	S = \left( 
 	\begin{array}{cccc}
 		1  & 0  &  \hdots & 0 \\
 		1  & 1  &  \hdots & 0 \\
 		\vdots  & \vdots  &  \ddots & \vdots \\
 		1  & 1  &  \hdots & 1 \\
 	\end{array}	
 	\right), S^{-1} = \left( 
 	\begin{array}{ccccc}
 		1  & 0  &  \hdots   & 0 \\
 		-1  & 1  &  \hdots  & 0 \\
 		\vdots  &   \ddots &  \ddots & \vdots \\
 		0  &   \hdots & -1  & 1 \\
 	\end{array}	
 	\right)
 \end{align} 
We denote the prediction horizon of MPC as $T$. Then, the MPC objective for controlling a platoon of AVs is defined as,
\begin{align} \label{eq_min}
		\begin{aligned}
		\displaystyle
		\min~ & J(u(k),\dots,u(k+T-1))= J_1 + J_2 + J_3\\
		\text{s.t.} ~~&  \text{Eqs.}~\eqref{eq_constv},\eqref{eq_consta},\eqref{eq_constp}
	\end{aligned}
\end{align}
where
\begin{align} \label{eq_J1}
	J_1 &:= \frac{1}{2} \sum_{m=1}^T \tau^2 u^\top(k+m-1) S^{-\top} Q_{u,m} S^{-1} u(k+m-1)\\ \label{eq_J2}
	J_2 &:= \frac{1}{2} \sum_{m=1}^Te_p^{\top}(k+m) Q_{p,m} e_p(k+m) \\ \label{eq_J3}
	J_3 &:= \frac{1}{2} \sum_{m=1}^T e_v^{\top}(k+m) Q_{v,m} e_v(k+m)
\end{align}
with positive semidefinite weight matrices $Q_{u,m},Q_{p,m},Q_{v,m}$  to adjust the control objective. 
Each of these cost functions is associated with a specific objective of the MPC. The term $J_1$ penalizes the input term and defines the ride comfort. The terms $J_2$ and $J_3$ penalize the position and velocity errors respectively and define the traffic stability and smoothness of the AV platooning. Then, for each $m=1,\dots,T$ and any fixed time-instant $k$ the followings hold:
\begin{align}
	v_i(k+m) &= v_i(k) + \tau \sum_{h=0}^{m-1} u_i(k+h) \\
	e_{v}(k+m) &= e_v(k) + \tau \sum_{h=0}^{m-1} e_u(k+h) \\ \nonumber
	e_{p}(k+m) &= e_p(k) + m \tau e_v(k) \\&+ \tau^2 \sum_{h=0}^{m-1} \frac{2(m-h)-1}{2} e_u(k+h) \\
	e_{u}(k+m) &= S^{-1}(u_0(k)\mb{1}_n - u(k+m)) 	
\end{align}
One can reformulate the MPC objective as a quadratic optimization problem with convex constraints in the following form:
\begin{align} \label{eq_min1}
 	\begin{aligned}
 		\displaystyle
 		\min~ & J(y)= \frac{1}{2} y^\top \Omega y + c^\top y + d \\
 		\text{s.t.} ~~&  y_i \in \mc{X}_i, (H_i(y))_m \leq 0
 	\end{aligned}
 \end{align}
for all $i=1,\dots,n$ and $m=1,\dots,T$ with $y = (y_1,\dots,y_n) \in \mathbb{R}^{nT}$, $y_i = (u_i(k),\dots,u_i(k+T-1)) \in \mathbb{R}^{T}$, $\Omega$ as a positive definite matrix (which is defined later). The constraints are defined as 
\begin{align} \nonumber
	 \mc{X}_i :=& \{z|a_{\min}\mb{1}_n \leq z \leq a_{\max}\mb{1}_n\} \\ \label{eq_Xi} &\cup \{z|(v_{\min}-v_i(k))\mb{1}_n \leq \tau S z \leq (v_{\max}-v_i(k))\mb{1}_n \}
\end{align}
which is convex and represents a polyhedral set. The other convex constraint is

\small \begin{align} \nonumber
	(H_i&(y_{i-1},y_i))_m := -\Big(p_{i-1}(k)+m\tau v_{i-1}(k) - p_{i}(k)-m\tau v_{i}(k)\Big) \\ \nonumber
	&-\tau^2 \sum_{h=0}^{m-1} \frac{2(m-h)-1}{2}(u_{i-1}(k+m)-u_{i}(k+m)) +l\\ \nonumber
	&+\epsilon v_i(k) + \epsilon \tau \sum_{h=0}^{m-1} u_i(k+m) - \frac{1}{2 a_{\min}}\Big(\tau^2 (\sum_{h=0}^{m-1} u_i(k+m))^2 \\ \label{eq_hi} &+2 \tau (v_i(k)-v_{\min})\sum_{h=0}^{m-1} u_i(k+m)+(v_i(k)-v_{\min})^2\Big) \leq 0
\end{align}\normalsize
For the closed-form expression of the objective matrix, define $\Omega = E^\top \Lambda E$ with $nT\times nT$ matrix 

\small \begin{align} \label{eq_lambda}
 \Lambda = \left( 
	\begin{array}{cccc}
		\Lambda_{1,1}+\tau^2 \tilde{Q}_{u,1}  & \Lambda_{1,2}  &  \hdots   & \Lambda_{1,T} \\
		\Lambda_{2,1}  & \Lambda_{2,2} +\tau^2 \tilde{Q}_{u,2}  &  \hdots   & \Lambda_{2,T} \\
		\vdots  &   \vdots &  \ddots & \vdots \\
		\Lambda_{T,1}  & \Lambda_{T,2}   &  \hdots   & \Lambda_{T,T} +\tau^2 \tilde{Q}_{u,T}
	\end{array}	
	\right)
\end{align} 
\begin{align}\nonumber
	\Lambda_{i,j}:=&S^{-\top}\Big(\sum^T_{\max (i,j)} \big(\frac{\tau^4}{4}(2(m-i)+1)(2(m-j)+1)Q_{p,m} \\&+ \tau^2 Q_{v,m}  \big) \Big)S^{-1} \in \mathbb{R}^{n\times n}
\end{align}\normalsize
and permutation matrix $E$ with entries defined as
\begin{align}\nonumber
		E_{ij} = \left\{ \begin{array}{ll}
			1, & \text{if}~  i=nk+m, j = T(m-1)+k+1\\
			0, & \text{otherwise}.
		\end{array}\right.
\end{align}
Similarly, the closed form expression for the linear term  in \eqref{eq_min1} is
\begin{align} \nonumber
	c^\top y +d = &-\sum_{i=1}^T\Big(\sum_{i=m}^T(\frac{\tau^2}{2}(2(m-i)+1) \gamma^\top(k)Q_{p,m} \\&+\tau \zeta^\top(k) Q_{v,m}) \Big) S^{-1}  u(k+i-1)
\end{align}
with $\gamma(k) := e_p(k)+m\tau e_v(k)+\tau^2 \sum_{j=0}^{m-1} \frac{2(m-j)-1}{2} S^{-1} \mb{1}_n u_0(k)$, $\zeta(k) := e_v(k) +\tau^2 \sum_{j=0}^{m-1}  S^{-1} \mb{1}_n u_0(k)$.

Note that every cost term in optimization \eqref{eq_min1} is associated with an AV $i$ and its neighbouring AVs $j \in \mc{N}_i$. In this direction define $\widehat{y}_i := (y_i,(y_{i,j})_{j \in \mc{N}_i})$ where the new variable $y_{i,j}$ denotes the predicted values for $y_j$ associated to the neighboring AVs $j \in \mc{N}_i$. In the light of coordination among the network of AVs, define the consensus subspace over the network $\mc{G}_W$ as,
\begin{align} \label{eq_cons_const}
	\mc{C} := \{\widehat{y}| y_{i,j} = y_j, \text{for all}~(i,j) \in \mc{E}\}
\end{align}
Other than this constraint, the other constraints in \eqref{eq_min1} can be added to the objective as a penalty term \cite{bertsekas1975necessary,nesterov1998introductory}. This eliminates these constraints via extra proper nonlinear barrier functions in the objective function. For example, any constraint in the form $y_{\min} \leq y_i \leq y_{\max}$ can be penalized by $f_i^{\sigma}(y_i) = \lambda ([y_i - y_{\max}]^+ + [y_{\min} - y_i ]^+)$, with the function $[x]^+ :=\max \{x, 0\}^\sigma,~\sigma \in \mathbb{N}$, and the parameter $ \lambda >0$ as the penalty coefficient that determines the weight of the constraint feasibility in the optimization process. Using such penalty terms, one can modify the objective function in \eqref{eq_min1} as,
\begin{align} \label{eq_min2}
	\begin{aligned}
		\displaystyle
		\min~ & F(\widehat{y})= \sum_{i=1}^n F_i(\widehat{y}_i) = \sum_{i=1}^n J_i(\widehat{y}_i) + \widetilde{F}_{\mc{X}_i}(\widehat{y}_i) + \widetilde{F}_{H_i}(\widehat{y}_i) \\
		\text{s.t.} ~~&  \widehat{y}_i \in \mc{C}
	\end{aligned}
\end{align}
with the objective variable $\widehat{y} = (\widehat{y}_1,\dots, \widehat{y}_n) \in \mathbb{R}^p$ and global extended-valued MPC objective $F: \mathbb{R}^p \rightarrow \mathbb{R}$ including the penalty terms $\widetilde{F}_{\mc{X}_i}(\widehat{y}_i)$ and $\widetilde{F}_{H_i}(\widehat{y}_i)$ addressing the constraints given by Eqs. \eqref{eq_Xi} and \eqref{eq_hi}. The formulation \eqref{eq_min2} represents a convex optimization problem with consensus-constraint given by Eq.~\eqref{eq_cons_const} that can be solved via fully distributed algorithms as discussed in the next section.

\section{Distributed Algorithm for MPC Optimization} \label{sec_dist}
Distributed algorithms are widely used to solve cooperative optimization problems over sensor networks and multi-agent systems. In this section, we provide a fully distributed algorithm to solve the MPC optimization problem \eqref{eq_min2}. This algorithm allows every vehicle to locally optimize its MPC objective by sharing relevant information over the communication network $\mc{G}_W$. Over this communication network, every vehicle is represented by a computing node and performs local data processing to iteratively solve the optimization problem.

The proposed algorithm introduces a new gradient-tracking (GT) variable $z_i$ at each computing node (vehicle) to follow the gradient of the local objective function $F_i$. The algorithm requires the weight factors at the communication links to be positive and balanced, i.e.,
\begin{align}
	&W=[w_{ij}], \left\{
	\begin{array}{ll}
		w_{ij}>0, & \text{If}~ j \in \mc{N}_i \\
		0, & \text{Otherwise}.
	\end{array}\right. \\ \label{eq_wb}
&\sum_{j=1}^{n} w_{ij}=\sum_{i=1}^{n} w_{ij}
\end{align}   
Every entry $w_{ij}$ at AV node $i$ denotes the weight factor on the information received from neighbouring AVs $j \in \mc{N}_i$. Define $\overline{W}$ as the Laplacian matrix associated with $W$ as
\begin{align}
	\overline{W}=[\overline{w}_{ij}], \left\{
	\begin{array}{ll}
		-\sum_{i=1}^{n} w_{ij}, & i=j \\
		w_{ij}, & i\neq j.
	\end{array}\right.
\end{align}
Given that the cooperation network of AVs is connected, the eigenvalues of $\overline{W}$ are all in the left-half-plane (towards stability) with one isolated zero eigenvalue \cite{olfatisaberfaxmurray07,olfati2004consensus}. The eigenspectrum of this Laplacian matrix plays a key role in the stability of the distributed optimization algorithm \cite{ddsvm}. The absolute value of the second largest eigenvalue of $\overline{W}$, denoted by $|\lambda^W_2|$,
is known as the algebraic connectivity\cite{de2007old} and determines the convergence rate of the optimization mechanism.

Given the graph-theoretic background, the main algorithm to solve the optimization problem \eqref{eq_min2} in a distributed way is provided as follows:
\begin{align} \label{eq_xdot_g}	
	\dot{{\widehat{y}}}_i &= \sum_{j \in \mc{N}_i} w_{ij} (q(\widehat{y}_j)-q(\widehat{y}_i))-\alpha z_i, \\ \label{eq_ydot_g}
	\dot{z}_i &= \sum_{j \in \mc{N}_i} w_{ij} (q(z_j)-q(z_i) ) + \partial_t \nabla F_i(\widehat{y}_i),
\end{align}
where $\partial_t$ denotes the time-derivative, $\alpha$ as the GT step rate, and function $q(\cdot)$ denotes the logarithmic quantization defined as
\begin{align}\label{eq_hl_q}
	q(x) = \mbox{sgn}(x)\exp\left(\rho\left[\dfrac{\log(|x|)}{\rho}\right] \right),
\end{align}
with $[\cdot]$ as rounding to the nearest integer, $\mbox{sgn}(\cdot)$ as the sign function, $\mbox{exp}(\cdot)$ as the exponential function, and $\mbox{log}(\cdot)$ as the (natural) logarithm function. The parameter  $\rho$ denotes the quantization level. Note that in the proposed solution we take into account possible data quantization on the network of AVs, which is not considered in many existing literature \cite{shen2022fully,shen2022nonconvex,qureshi2020s,nedic2014distributed}. Recall that quantization is a common approach in data-sharing networks that reduces the amount of data to be transmitted in resource-constrained environments.  By reducing the size of the data packets, quantization can lead to faster transmission times. This is particularly important in real-time applications like intelligent transportation systems, where timely data sharing can enhance decision-making and improve safety \cite{gholami2022survey}.

The discrete-time version of the continuous-time dynamics \eqref{eq_xdot_g}-\eqref{eq_ydot_g} is as follows:

	\small
	\begin{align} \label{eq_xdot_gd}	
		{\widehat{y}}^{t+1}_i &= {\widehat{y}}^{t}_i +\sum_{j \in \mc{N}_i} w_{ij} (q(\widehat{y}_j^t)-q(\widehat{y}_i^t))-\widetilde{\alpha} z^{t}_i, \\ \label{eq_ydot_gd}
		{z}^{t+1}_i &= {z}^{t}_i +\sum_{j \in \mc{N}_i} w_{ij} (q(z^{t}_j)-q(z^{t}_i) ) + \nabla F_i({\widehat{y}}^{t+1}_i)-\nabla F_i({\widehat{y}}^{t}_i),
	\end{align} \normalsize 
To present the GT nature of the proposed dynamics, sum all the AV computing states over the network. Recall from Eq. \eqref{eq_wb} that the consensus matrix $W$ is weight-balanced and we have $\sum_{i=1}^n \sum_{j \in \mc{N}_i} w_{ij} (q(z^{t}_j)-q(z^{t}_i) ) = 0$ and $\sum_{i=1}^n \sum_{j \in \mc{N}_i} w_{ij} (q(z^{t}_j)-q(z^{t}_i) )=0$. Therefore, we get
\begin{align}  \label{eq_sumydot}
	\sum_{i=1}^n {z}_i^{t+1} 
	&= \sum_{i=1}^n {z}_i^{k} + \sum_{i=1}^n \nabla F_i(\widehat{y}^{t+1}_i)-\sum_{i=1}^n \nabla F_i(\widehat{y}^{t}_i), \\
	\label{eq_sumxdot}
	\sum_{i=1}^n {\widehat{y}}_i^{t+1} 
	&= \sum_{i=1}^n {\widehat{y}}_i^{t}-\widetilde{\alpha} \sum_{i=1}^n {z}_i^{t}.
\end{align}
Then, by setting the initial condition for the auxiliary variable as~$z_i(0)=\mb{0}_{p}$, we get the following: 

\small \begin{align} \label{eq_sumxdot2}
	\sum_{i=1}^n {\widehat{y}}_i^{t+1} = -\widetilde{\alpha} \sum_{i=1}^n {z}_i^{t+1} = -\widetilde{\alpha} \sum_{i=1}^n \nabla F_i({\widehat{y}}_i^{t+1})= -\widetilde{\alpha} \nabla F({\widehat{y}}^{t+1}),
\end{align} \normalsize
The above equation shows how the sum of the auxiliary variable tracks the gradient of the MPC objective function. Our proposed distributed solution is summarized in Algorithm~\ref{alg_1}.
\begin{algorithm} 
	\textbf{Data:}  $F(\widehat{y})$, $\mc{G}_W$, $W$, $\widetilde{\alpha}$  \\	
	\textbf{Initialization:} ${z}_i(0)=\mb{0}_{p}$, random ${\widehat{y}}_i(0)$
	\\
	\For{$t=0,1,2,\dots$}{
		AV $i$ receives $q(\widehat{y}_j^t)$ and $q(z^{t}_j)$ from neighboring AVs $j \in \mc{N}_i$\;
		$\widehat{y}_i^{t+1} \leftarrow \widehat{y}_i^{t} +\sum_{j=1}^{n} w_{ij} (q(\widehat{y}_j^{t})-q(\widehat{y}_i^{t}))-\widetilde{\alpha} z^{t}_i$\;
		$b_i^t \leftarrow \nabla F_i(\widehat{y}_i^{t+1})-\nabla F_i(\widehat{y}_i^{t})$\;
		$z^{t+1}_i \leftarrow z^{t}_i +\sum_{j=1}^{n} w_{ij} (q(z^{t}_j)-q(z^{t}_i) ) + b_i^t$\;
		AV $i$ shares $\widehat{y}_i^{t+1}$ and $z^{t+1}_i$ over $\mc{G}_W$\;
	}
	\textbf{Return:}  optimal values $\widehat{y}^*$ and $F^*$\;	
	\caption{Distributed MPC optimization at each AV $i$. } \label{alg_1}
\end{algorithm} 

The convergence proof of this algorithm is based on matrix perturbation theory \cite{bhatia2007perturbation}, algebraic graph theory and eigenspectrum analysis \cite{godsil}. Here, we provide the sketch of the proof and more detailed proof can be found in \cite{tnse24,ddsvm} for general sector-bound nonlinearities on data-sharing networks. The terms including $\widetilde{\alpha}$ can be considered as a perturbation to the main consensus-based dynamics. Recall from consensus literature that all the eigenvalues of the Laplacian consensus matrix $\overline{W}$ are in the left-half-plane except one isolated zero eigenvalue \cite{olfati2004consensus}. This guarantees consensus-based stability. In the given dynamics \eqref{eq_xdot_gd}-\eqref{eq_ydot_gd}, the main system matrix includes two sets of consensus matrices $\overline{W}$ in its (block) diagonals each associated with one of the system parameters $\widehat{y}$ and $z$. Therefore, the main system dynamics is associated with two zero eigenvalues. By adding perturbation associated with the parameter $\widetilde{\alpha}$ one set moves to the left-half-plane for sufficiently small $\widetilde{\alpha}$, ensuring stability. The other one remains zero, ensuring consensus among all the computing nodes. Recall from \cite{ddsvm} that the sufficiency bound for convergence is defined as $\widetilde{\alpha} \leq \frac{|\lambda_2|}{\eta}$ with $\eta$ as the Lipschitz bound on the gradient of the objective function \eqref{eq_min2}. To improve the convergence rate of the algorithm one may add extra signum-based non-Lipschitz dynamics \cite{spl24} or additive momentum terms \cite{nguyen2023accelerated,xin2019distributed}.

\begin{rem}
	The proposed optimization dynamics converges in the presence of log-scale quantization over the communication network of AVs. Recall that most existing distributed optimization algorithms assume ideal communication channels over the network and ignore quantization constraints in existing real-world resource-constrained communication devices. Therefore, these algorithms may result in optimality gaps in real-world quantized setups, while the proposed solution in this paper reaches exact convergence. This is better illustrated in the simulation section. 
\end{rem}

\section{Simulations} \label{sec_sim}
\begin{figure}[hbpt!]
	\centering
	\includegraphics[width=2.5in]{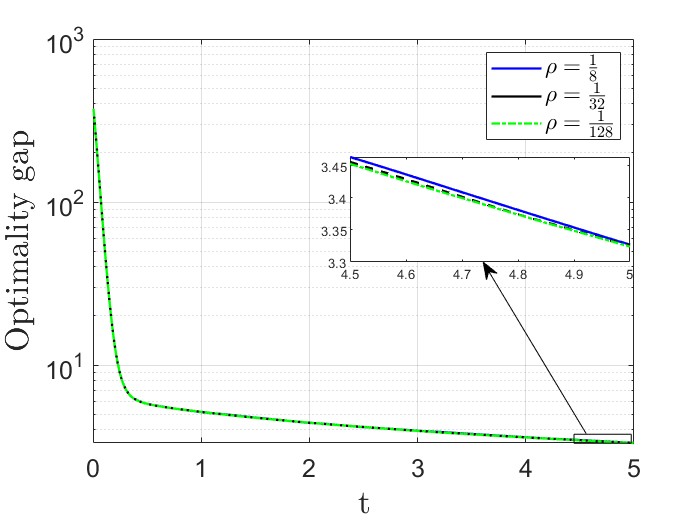}
	\caption{This figure shows the iterative distributed optimization of the MPC cost function~\eqref{eq_min2} for different log-scale quantization levels.  }
	\label{fig_rho}
\end{figure}

In this section, we provide simulations to verify the convergence of the proposed Algorithm~\ref{alg_1} solving the MPC optimization problem \eqref{eq_min2} over time-horizon $T=5$. We run our distributed algorithm on MATLAB and on a laptop with the following features:  Intel(R) Quad-Core i5 CPU 1.80GHz
and 8GB RAM.
For simulation, we consider a cyclic network of $n=10$ AVs under the MPC setup in Section~\ref{sec_prob} subject to the log-scale quantized data-sharing network with different quantization levels $\rho$. We evaluate
the numerical accuracy of distributed Algorithm~\ref{alg_1} by simulating the optimality gap between the
proposed solution and the centralized scheme solved by MATLAB CVX toolbox. For simulation, we set random entries for quadratic cost function \eqref{eq_min2}, and for the penalty terms to address the constraints we set $\sigma = 2$ and $\lambda=1$. The simulation results for different values of quantization level $\rho$ are shown in Fig.~\ref{fig_rho}.
As it is clear from the figure, the cost function is decreasing over time subject to different quantization levels $\rho=\frac{1}{8},\frac{1}{32},\frac{1}{128}$.

\begin{figure}[hbpt!]
	\centering
	\includegraphics[width=2.5in]{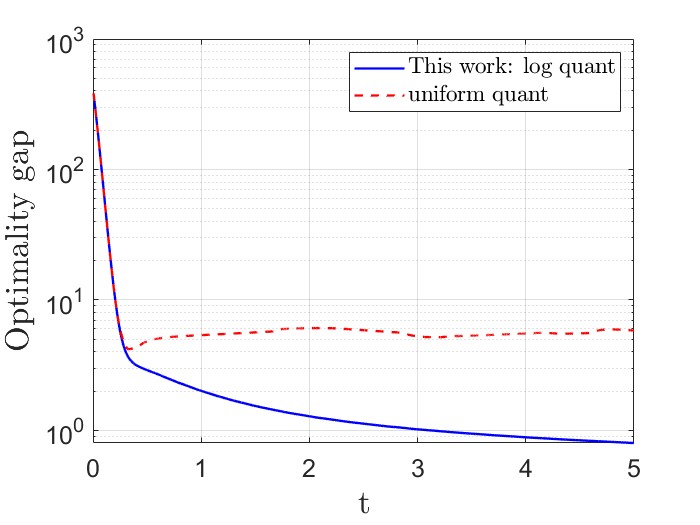}
	\caption{This figure compares the iterative distributed optimization of the MPC cost function~\eqref{eq_min2} under log-scale quantization versus uniform quantization. The algorithm under uniform quantization results in a large optimality gap, while the log-quantized distributed algorithm in this work converges toward the optimal value. }
	\label{fig_log_uni}
\end{figure}

Next, we compare our log-quantized optimization scheme with other existing distributed optimization methods subject to uniform quantization, for example \cite{rikos2024distributed}. Recall that
in uniform quantization, the range of input values is divided into equal-sized intervals and is defined as \cite{liu2022nonuniform},
\begin{align}\label{eq_hl_qu}
	q_u(x) = \rho\left[\dfrac{x}{\rho}\right],
\end{align}
with $\rho$ as the uniform quantization level. For simulation, we set both uniform and logarithmic quantization levels as $\rho= 0.0625$. The simulation results are shown in Fig.~\ref{fig_log_uni}.
As it is clear from the figure, distributed optimization under uniform quantization results in a large optimality gap. On the other hand, since logarithmic quantization is a sector-bound nonlinearity \cite{fu2005sector}, the optimality gap of the proposed algorithm in this paper converges toward zero. This is because log-quantization adapts to the varying scales of the gradient parameters near the optimal point. 

\section{Conclusion and Future Works} \label{sec_con}
This paper develops a fully distributed algorithm for MPC-based platooning optimization while considering log-quantized data exchange among the connected vehicles. By modelling the MPC framework as a cooperative optimization formulation, each vehicle locally solves its objective function using quantized information received from its neighbouring vehicles. Using a log-quantization setup, we show that the proposed algorithm reaches optimal convergence while a uniform quantization setup may lead to some optimality gap. 
  
As a future research direction, application to quantized communication over large-scale IoT systems \cite{iot} is of interest. The other factor that may affect the optimization performance is latency in the communication network of vehicles. Distributed optimization in the presence of time-delays \cite{cdc22} is another future research direction.
 
\bibliographystyle{IEEEbib}
\bibliography{bibliography}

\end{document}